\documentstyle[mnextra]{mn}
\input{epsf}
\seceqnum           
\def\etal{{\rm et al.} }

\def\Mpc  {{\it h}^{-1}\, {\rm Mpc}}

\def\Msol {{\it h}^{-1}\, {\rm M_\odot}}
\def\d    {{ \rm d}}
\def\ie {{\rm i.e. }}
\def\eg {{\rm e.g. }}
\def\log {{\rm log}}
\def\dcz  {\delta_{\rm c}(z)}
\def\lsim{\mathrel{\hbox{\rlap{\hbox{\lower4pt\hbox{$\sim$}}}\hbox{$<$}}}}
\def\gsim{\mathrel{\hbox{\rlap{\hbox{\lower4pt\hbox{$\sim$}}}\hbox{$>$}}}}

\begin{document}
\title[$\Omega_0$ from cluster evolution]
{Measuring $\Omega_0$ using cluster evolution}
\author[V.R. Eke et al]{Vincent R. Eke${}^1$, Shaun Cole${}^2$, Carlos S. 
Frenk${}^2$, and J. Patrick Henry${}^{3}$ \\
$^1${Institute of Astronomy, University of Cambridge, Madingley Road, 
Cambridge CB3 OHA}\\
$^2${Department of Physics, University of Durham, South Road, Durham DH1 3LE}\\
$^3${Institute for Astronomy, 2680 Woodlawn Drive, Honolulu, HI 96822}
}

\maketitle

\begin{abstract}

The evolution of the abundance of galaxy clusters depends sensitively on
the value of the cosmological density parameter, $\Omega_0$. Recent 
ASCA data are used to quantify this evolution as measured by the cluster X-ray
temperature function. A $\chi^2$ minimisation fit to the cumulative
temperature function, as well as a maximum likelihood estimate (which
requires additional assumptions about cluster luminosities), lead to the
estimate $\Omega_0\approx 0.45 \pm 0.2$ (1-$\sigma$ statistical error).
Various systematic uncertainties are considered, none of which enhance
significantly the probability that $\Omega_0=1$. These conclusions hold for
models with or without a cosmological constant, \ie with $\Lambda_0=0$ or
$1-\Omega_0$. The statistical uncertainties are at least as large as
any of the individual systematic errors that have been considered here,
suggesting that additional temperature measurements of distant clusters
will allow an improvement in this estimate. An alternative method that 
uses the highest redshift clusters to place an upper limit on $\Omega_0$
is also presented and tentatively applied, with the result that
$\Omega_0=1$ can be ruled out at the $98$ per cent confidence level.
Whilst this method does not require a well-defined statistical sample of
distant clusters, there are still modelling uncertainties that preclude
a firmer conclusion at this time.

\end{abstract}

\section{Introduction}

Galaxy clusters are the largest virialised objects in the universe and, as
such, provide useful cosmological probes. For example, the evolution in the
abundance of clusters is strongly dependent on, and is therefore a sensitive
probe of, the cosmological density parameter, $\Omega_0$, (\eg Evrard 1989;
Frenk \etal 1990; Lilje 1992; Oukbir \& Blanchard 1992).

The amount of observational data concerning high-redshift cluster properties 
has increased significantly in recent years. 
{\it EMSS} (Henry \etal 1992; Gioia \& Luppino 1994), {\it ASCA} 
(\eg Donahue 1996; Mushotzky \& Scharf 1997; Henry 1997) and
{\it ROSAT} (Ebeling \etal 1997, Burke \etal 1997, Rosati \etal 1998) 
measurements of the X-ray emitting intracluster 
plasma now complement the low-redshift studies carried out by Edge \etal 
(1990) and David \etal (1993). 
In addition, the {\it CNOC} survey provides
galaxy velocity dispersions for a well-defined sample of high-redshift
clusters (Carlberg \etal 1996) and further information on mass distributions is
coming from weak gravitational lensing measurements (\eg Luppino \& Kaiser 
1997; Smail \etal 1997).

The modelling of the cluster population has relied heavily upon the
expression provided by Press \& Schechter (1974) for the mass function of
haloes. This has received support from dark matter simulations of
structure formation (\eg Efstathiou \etal 1988; White, Efstathiou \& Frenk
1993, Lacey \& Cole 1994). More recently the treatment of a baryonic
component in simulations has led to a closer link with observational
measurements (Evrard 1990; Cen \& Ostriker 1994; Metzler \& Evrard 1994;
Bryan \etal 1994; Navarro, Frenk \& White 1995, hereafter NFW). In
addition, analytical work investigating the evolution of properties of the
cluster population as a whole has been performed by Kaiser (1986, 1991),
Evrard \& Henry (1991) and Bower (1997).

There is now sufficient overlap between these strands of research to allow
useful conclusions concerning cosmological parameters to be drawn from the
available data (see however Colafrancesco, Mazzotta \& Vittorio 1997).
Several groups have already performed this operation (Henry 1997; Carlberg
\etal 1997b; Bahcall, Fan \& Cen 1997; Sadat, Blanchard \& Oukbir 1998;
Blanchard \& Bartlett 1997; Reichart \etal 1998), although the lack of 
consensus on the
measured value for $\Omega_0$ casts doubt on the reliability of at least
some of these studies. The purpose of this paper is to investigate the
systematic uncertainties in the estimation of $\Omega_0$ using cluster
evolution, and to understand why the various studies differ. 

In Section~\ref{sec:model} the cosmology-dependent model for describing the 
evolution of
the cluster population is described. The procedure for and results from
fitting the models to
the observed temperature functions are presented in Section~\ref{sec:evol}.
Section~\ref{sec:maxl} contains
a description of a maximum likelihood method used to estimate cosmological 
parameters using the cluster redshifts and temperatures. The results from
these two methods are compared with other studies in Section~\ref{sec:others}.
In Section~\ref{sec:order}
an alternative method is explored for using high-redshift cluster data to 
constrain $\Omega_0$. A discussion of the different techniques is presented
in Section~\ref{sec:tech}. The implications of these results for $\Omega_0=1$
scenarios are discussed in Section~\ref{sec:omnotone}, and conclusions are
summarized in Section~\ref{sec:conc}.

\section{Model}\label{sec:model}

The model of the population of hot galaxy clusters employed here is the 
same one described by Eke, Cole \& Frenk (1996). 
This is based upon the analytical
expression for the comoving number density of dark matter haloes first 
derived by Press \& Schechter (1974):
\begin{equation}
\frac{\d n}{\d M_{\rm vir}} = \left( \frac{2}{\pi} \right)^{\frac{1}{2}}
\frac{\bar \rho}{M_{\rm vir}^2}
\frac{\dcz}{ \sigma}
\left\vert \frac{\d \ln \sigma} {\d \ln M_{\rm vir}}\right\vert
\exp \left[- \frac{\dcz^2}{2\sigma^2} \right].
\label{ps}
\end{equation}
Here $M_{\rm vir}$ denotes the halo virial mass, defined such that a halo
has a mean density $\Delta_{\rm c}$ times the critical density at redshift
$z$, where $\Delta_{\rm c}$ is evaluated using the spherical collapse
model; $\bar \rho$ represents the present-day mean density of the universe
and $\sigma(M_{\rm vir})$ the present-day, linear theory rms density
fluctuation in spheres containing a mean mass $M_{\rm vir}$. Throughout
this paper, we use the fit to the cold dark matter power spectrum of
Bardeen \etal (1986). The shape of the power spectrum and its amplitude at
$8 \Mpc$ are described by $\sigma_8$ and $\Gamma$ respectively
(Efstathiou, Bond \& White 1992). The spherical collapse threshold,
$\delta_c$, is employed for most of the analysis in the rest of this paper. 
Bryan \& Norman (1998)
report that their work, combined with the majority of published results,
suggests that a slightly ($< 10$ per cent) higher value provides a better
fit to mass functions from dark matter simulations. However, the recent
simulations of Tozzi \& Governato (1997) appear to support a lower value.
This will be considered further in Section~\ref{sec:omnotone}. Monaco
(1998) has provided a review of the present understanding of the mass
function. 

An additional crucial ingredient in the model is the conversion
from virial mass to emission-weighted X-ray temperature:
\begin{eqnarray}
kT_{\rm gas} &=& \frac{7.75}{\beta_{\rm TM}} \left(\frac{6.8}{5X+3}\right)
\left(\frac{M_{\rm vir}}{10^{15}\Msol}\right)^{\frac{2}{3}}
 \cr
&& \times (1+z)
\left(\frac{\Omega_0}{\Omega(z)}\right)^{\frac{1}{3}}
\left(\frac{\Delta_{\rm c}}{178}\right)^{\frac{1}{3}} \,{\rm keV}.
\label{kev}
\end{eqnarray}
In this equation, $\beta_{\rm TM}$ is a conversion factor, the numerical value 
of which will be discussed shortly, and $X$ is the
hydrogen mass fraction which is taken to be $X=0.76$. 
If the specific galaxy kinetic energy equals the specific gas thermal
energy within the virial radius, then a value of $\beta_{\rm TM}=1$ would
be appropriate for an isothermal gas. Alternatively, this value follows
from applying hydrostatic equilibrium at the virial radius and assuming
both that the sum of the logarithmic derivatives of the gas density and
temperature profiles at this radius is $-2$ and that the gas temperature
at this radius equals the overall emission-weighted value. Neither of
these reasons is particularly persuasive. Firstly, there is likely to be
some residual kinetic energy in the gas as a result of incomplete
thermalisation (Evrard 1989). Secondly, numerical simulations 
show that the gas density is dropping off faster than $r^{-2}$ at
the virial radius and that the temperature is also decreasing rather than
remaining constant (\eg NFW). The value of $\beta_{\rm TM}$ chosen here is 
based solely upon the results of hydrodynamical cluster simulations. These
usually include shock heating but no radiative cooling or heating from any
source. Such experiments (NFW; Evrard, Metzler \& Navarro 1996; Eke,
Navarro \& Frenk 1998; Bryan \& Norman 1998; Frenk \etal 1998) suggest
that $\beta_{\rm TM}=1.0$ is a reasonable choice, and it is adopted as the
default value in this paper, although the dependence of the results on
this assumption will be investigated in Section~\ref{sec:maxl}. Values
returned from different types of hydrodynamical simulations for a variety
of cosmologies range between $1$ and $1.3$ (see Bryan \& Norman 1998),
with a slight preference for larger values of $\beta_{\rm TM}$ when
$\Omega_0$ is larger. No significant mass dependence of $\beta_{\rm TM}$
is found.

\begin{figure}
\centering
\centerline{\epsfxsize=9.0cm \epsfbox{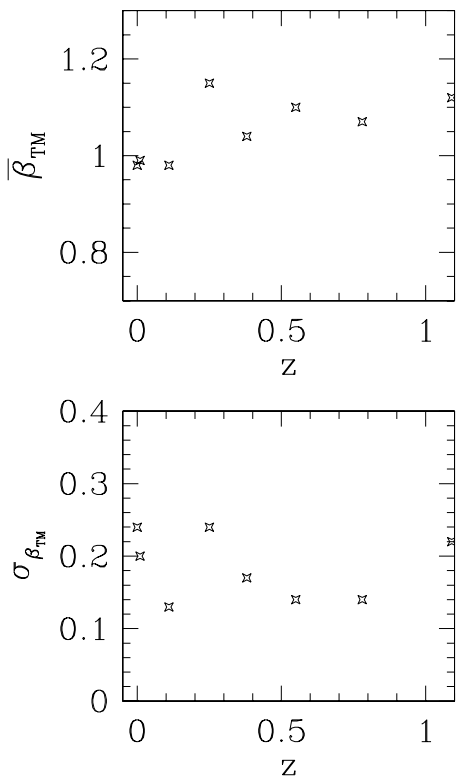}}
\caption{The top panel shows the evolution with redshift of the 
parameter $\beta_{TM}$,
relating the emission weighted X-ray temperature to the halo mass,
as measured from an ensemble of $10$ high-resolution hydrodynamical
simulations of an $\Omega_0=0.3$, $\Lambda_0=0.7$ cosmology (Eke \etal 1998).
The lower panel shows the standard deviation of the individual cluster 
$\beta_{TM}$ values about this mean value as a function of redshift.}
\label{fig:beta}
\end{figure}

The redshift dependence of 
the mean value of $\beta_{\rm TM}$ for the $10$ clusters simulated
by Eke \etal (1998), and the scatter among 
the individual cluster values are 
shown in Fig.~\ref{fig:beta}. Most importantly for estimating $\Omega_0$ from
cluster evolution out to $z\sim0.4$, there is no significant evolution
in either the mean $\beta_{\rm TM}$ or in the magnitude of the scatter 
about this value.
As the temperature function has such a steep dependence on cluster temperature,
uncertainties and possible biases in the mass-temperature conversion are very
important (\eg Pen 1998). 
A Gaussian scatter of $20$ per cent in the temperature at a given mass
has been included, consistent with Fig.~\ref{fig:beta}, 
when calculating the model predictions.
The neglect of physical processes associated with galaxy formation 
in hydrodynamical simulations of clusters is 
a concern, but the work of Metzler \& Evrard (1994) suggests that even extreme
feedback from stellar evolutionary effects cannot heat the intracluster 
medium of a rich cluster by more than about $15$ per cent. 
Similarly, NFW found that 
$\beta_{\rm TM}$ was decreased by about $20$ per cent if preheating was 
included in rich cluster simulations in order to match the slope of the
observed luminosity-temperature relation. The study by White (1991) 
comparing 
observed velocity dispersions with temperatures is also suggestive that 
non-gravitational processes have a relatively small effect on rich clusters.
In Section~\ref{sec:omnotone}, the possible extent to which the results
presented here might depend upon physics missing from the numerical simulations
will be considered further.

\section{Evolution of the cluster temperature function}\label{sec:evol}

\subsection{Estimating the cumulative temperature functions}\label{ssec:obs}

The cluster X-ray emission-weighted temperature measurements that are used here
come from two different sources. At low redshift, an updated version of 
the flux-limited ($f(2-10 $~keV$)> 3\times10^{-14}$Wm$^{-2}$
\footnote{For the benefit of older readers, $1$ W $\equiv 10^{7}$erg s$^{-1}$.}
) sample of $25$ clusters compiled by Henry \& 
Arnaud (1991, hereafter HA91) is employed (see Henry 1998 for details).
These authors report that this sample is ``at least 90 per cent complete''.
In this Section, a completeness of 100 per cent will be assumed, but the
effect of this uncertainty will be investigated further in 
Section~\ref{sec:maxl}. The average $1\sigma$ measurement uncertainty in 
these temperatures is $7$ per cent.
For the high redshift data, the $10$ {\it EMSS}~$0.3<z<0.4$ 
cluster temperatures 
obtained by Henry (1997) are employed. These have an average $1\sigma$
uncertainty of $14$ per cent. No systematic differences are found between 
these results and those of Mushotzky \& Scharf (1997) for the $5$ cases 
where they studied the same clusters.

In each case the cumulative cluster temperature function is estimated using
\begin{equation}
N(>kT) = \sum_{kT_i>kT} 1/V_{\rm max,{\it i}},
\label{esti}
\end{equation}
where $V_{\rm max,{\it i}}$ is the maximum volume in which cluster $i$ could be
detected given the survey selection criteria, and is calculated using the 
cluster fluxes and redshifts. 
A complication for the high-z sample is that the area of sky surveyed was a
function of the sensitivity, \ie only a small fraction of the total survey
area was observed at the lowest flux levels. Using the data in table~$3$ of 
Henry et al. (1992), the following expression has been found to give 
the area of sky, $A$, in which a cluster providing a detect cell flux,
$f_{\rm det}$, would have been detectable by the {\it EMSS}:
\begin{equation}
\frac{A(f_{\rm det})}{A_{\rm tot}}=
1-3.05~{\rm e}^{-0.41f_{\rm det}}+2.30~{\rm e}^{-0.77f_{\rm det}}.
\label{cover}
\end{equation}
Here, $f_{\rm det}$ is in units of $10^{-16}$W m$^{-2}$
and $A_{\rm tot}$ is the maximum area covered by the {\it EMSS} cluster
survey, $735$ square degrees. This fit is accurate to within $10$ per cent
at all detect cell fluxes above the Henry (1997) catalogue limit of 
$2.5 \times 10^{-16}$W m$^{-2}$.
Using the cluster detect cell fluxes and 
redshifts provided in table~$1$ of Henry (1997), the cluster flux can be
calculated as a function of redshift.
$V_{\rm max,{\it i}}$ is then evaluated by integrating the product
of $\d V/\d z$ and survey
area at a particular redshift from z=0.3 to z=0.4.

Before this estimate of $N(>kT)$ can be compared with the model, a
correction for incompleteness in the measured high-redshift temperature
function is required. This arises because of the non-zero scatter in the
luminosity-temperature relation and the fact that the high-redshift sample
has a non-zero lower redshift limit. As a consequence, at the low redshift
limit of the sample, the flux limit of $2.5 \times 10^{-16}$W m$^{-2}$
corresponds to a non-zero luminosity limit such that for any given
temperature, only clusters with luminosities higher than this limit will
be included. The incompleteness will be greater for cooler clusters
because, on average, they have lower luminosities. To calculate the
amplitude of this effect, the data of HA91 for low-redshift clusters were
used to provide both a mean luminosity as a function of cluster
temperature and a scatter about this mean. What observational evidence
there is suggests that the luminosity-temperature relation evolves very
little out to redshifts of $0.4$ (Tsuru \etal 1996; Mushotzky \& Scharf
1997; Henry 1997). This gave
\begin{equation}
{\rm log}_{10} \bar{L}_{37}^{0.3-3.5}=3.54{\rm log}_{10}(kT)-2.53,
\label{ltmean01}
\end{equation}
where $\bar{L}_{37}^{0.3-3.5}$ is the $0.3-3.5$~keV luminosity measured in 
$10^{37} h^{-2}$ W
and $kT$ is in units of keV, and a Gaussian scatter in log$_{10} L_{37}
^{0.3-3.5}$ of
\begin{equation}
\sigma = 0.69 - 0.47 {\rm log}_{10}(kT).
\label{ltscat01}
\end{equation}
At any given temperature, the fraction of clusters that are actually 
selected is then just given by
\begin{equation}
F(T,L_{\rm lim})=0.5~{\rm erfc}(x_{\rm lim}(T))
\end{equation}
where erfc represents the complementary error function, 
$x_{\rm lim}=(\log_{10} L_{\rm lim}-\log_{10} \bar{L})/(\sqrt{2}\sigma)$ 
and $L_{\rm lim}$ is the (cosmology-dependent) luminosity 
of a cluster at $z=0.3$ having a flux equal to $f_{\rm lim}$. To allow for the
fact that the catalogue flux limit refers to the {\it EMSS} detect cell whilst
the clusters are typically more extended, a factor of $2.1$ has been included
to account for the conversion between these two fluxes (see figure~$1$ of 
Henry \etal 1992).
\begin{figure}
\centering
\centerline{\epsfxsize=9.0cm \epsfbox{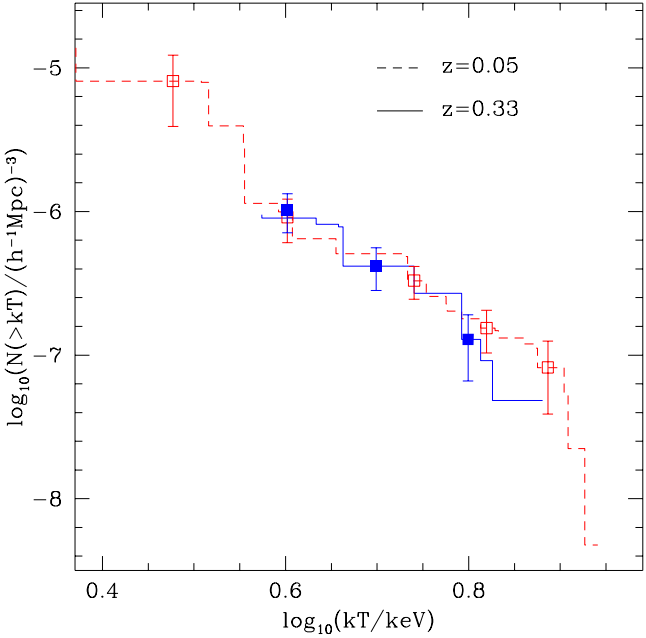}}
\caption{The stepped curves show the low- (dashed) and high- (solid) redshift
cumulative cluster temperature functions, calculated
assuming that $\Omega_0=1$. The 
points are the values that were used in the model fitting, and include the
incompleteness correction for the high-redshift case. Open and filled symbols
correspond to the low- and high-redshift data respectively.
Error bars were computed using bootstrap resampling.}
\label{fig:datinc}
\end{figure}
The product of this factor and the differential number density at the same
temperature can then be integrated with respect to temperature to give the 
correction to the cumulative temperature function. Even for temperatures 
as low as $4$~keV the incompleteness is no more than about $20$ per
cent. This can be seen in Figure~\ref{fig:datinc}, where the low redshift
cumulative cluster temperature function is shown as a stepped line along with 
that from the high-z sample, calculated
assuming that $\Omega_0=1$ and ignoring the
incompleteness correction. The points represent the data used for the 
quantitative comparison with the models, and they include the correction factor
which only has any significant effect on the lowest temperature measurement.
The $1\sigma$ uncertainties on the number densities come from a bootstrap
resampling procedure using $10^4$ catalogues.

\subsection{Comparison with model predictions}\label{ssec:comp}

\begin{figure}
\centering
\centerline{\epsfxsize=9.0cm \epsfbox{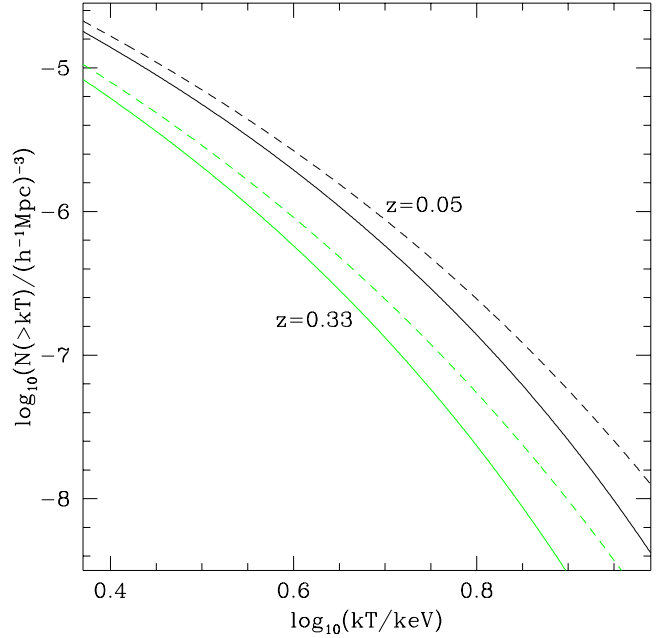}}
\caption{The predicted evolution of the cluster temperature function between 
$z=0.05$ and $z=0.33$ is shown for an $\Omega_0=1$ model with $\sigma_8=0.52$ 
and $\Gamma=0.25$, both before (solid) and after (dashed) a Gaussian 
smoothing with width $0.2T$ has been applied.}
\label{fig:models}
\end{figure}

Given that the cumulative cluster number density varies rapidly with 
temperature, uncertainties in either the mass-to-temperature conversion or 
in the 
observed temperatures of individual clusters will impact more on the results 
than any minor incompleteness. These effects can both be modelled
by smoothing the model temperature function before it is compared with
observational data. The width of the smoothing function, which has been
taken to be a Gaussian, has been estimated by adding the fractional 
uncertainties from the two sources in quadrature. From the results in
Fig.~\ref{fig:beta}, the scatter in individual cluster 
$\beta_{\rm TM}$ values is slightly less than $20$ per cent and it appears to 
be independent of redshift for $z<0.4$. While these simulations were of an
$\Omega_0=0.3$, $\Lambda_0=0.7$ model, in the absence of information to 
the contrary, this scatter is assumed to apply to other cosmological 
models as well. For both observational redshift ranges,
the temperature uncertainties are sufficiently small compared to the 
scatter in $\beta_{\rm TM}$, that neglecting them will
not affect the final results significantly. A
Gaussian smoothing width of $0.2T$ was therefore used. 
From Fig.~\ref{fig:models} it is apparent that this 
correction is not negligible compared with the amount of evolution that the 
$\Omega_0=1$ model predicts. However, if the width of the smoothing Gaussian
is similar for both the high- and low-redshift samples, then the amount of
evolution is largely unaffected because the cumulative temperature
function is almost a power law. The estimate of $\sigma_8$ will be decreased
slightly by this effect. Compared with the rapid evolution in the number of
clusters at a particular temperature, Fig.~\ref{fig:models} 
illustrates how little the
temperature at a fixed number density evolves, and how important any evolution
in the mass-temperature relation is over this range of redshifts.

The best-fitting parameter set ($\Omega_0$, $\Gamma$, $\sigma_8$) was
found by minimising $\chi^2$ for the models calculated at five and three
temperatures in the low- and high-redshift bins respectively. Bin
temperatures were chosen to be far apart, in order to reduce the
correlations between the different points. Nevertheless, the full
covariance matrix was employed when calculating the $\chi^2$ values. The
number of bins used was the minimum for which the results of the model
fitting were found to be robust to the exact positioning of the bins. If
any fewer bins are taken, then the results depend upon their chosen
temperatures. This is particularly so for the low-redshift data where the
observed function is less smooth than in the high-redshift case. The
statistical uncertainties in the best-fitting parameters were quantified 
using both the change in $\chi^2$ and the distribution of best-fitting
values from the $10^4$ bootstrap catalogues that were created from the
observational data.

\subsection{Results}\label{ssec:tfresults}

\begin{figure}
\centering
\centerline{\epsfxsize=9.0cm \epsfbox{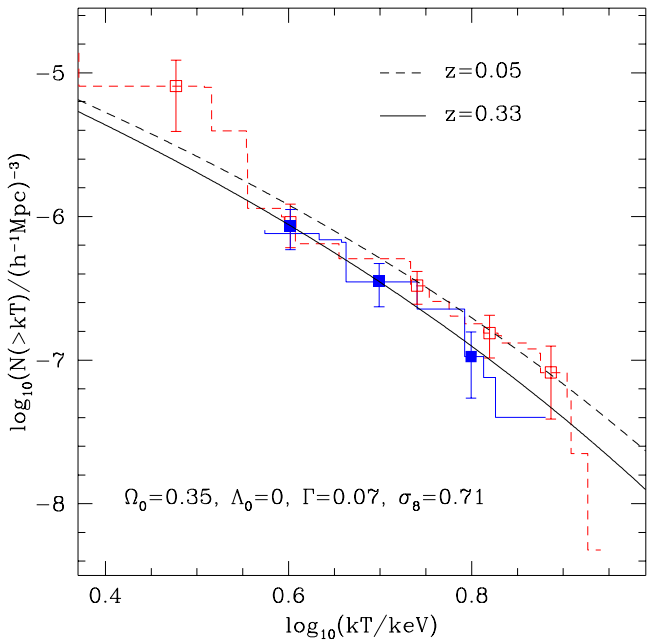}}
\caption{The best-fitting model and the corresponding data are shown for the
$\Lambda_0=0$ case. Low (high) redshift data
are represented with a dashed (solid) line and open (filled) squares as in
Fig.~\ref{fig:datinc}.}
\label{fig:bestfit}
\end{figure}

The $\chi^2$ minimisation gave rise to the following best-fitting parameter
values: $\Omega_0=0.35$, $\Gamma=0.07$ and $\sigma_8=0.71$ for
$\Lambda_0=0$ and $\Omega_0=0.31$, $\Gamma=0.07$ and $\sigma_8=0.78$ for
$\Lambda_0=1-\Omega_0$. In the latter case, two different effects act in
opposite directions on the most likely value of $\Omega_0$ compared to the
$\Lambda_0=0$ case. If $\Lambda_0$ is non-zero, then the expected amount
of evolution at a particular $\Omega_0$ increases because of the more
rapidly changing growth factor. However, the volume surveyed between
redshifts $0.3$ and $0.4$ also increases if a $\Lambda_0=1-\Omega_0$ term
is included, and this has the effect of decreasing the observed cluster
number density at high-redshift relative to the low-redshift measurement.
Given that the most likely $\Omega_0$ decreases, albeit only slightly,
when $\Lambda_0$ is assumed to be non-zero, the former effect is the more
important. The best-fitting curves for $\Lambda_0=0$, together with the
observational data points, are shown in Fig.~\ref{fig:bestfit}.

\begin{figure}
\centering
\centerline{\epsfxsize=8.5cm \epsfbox{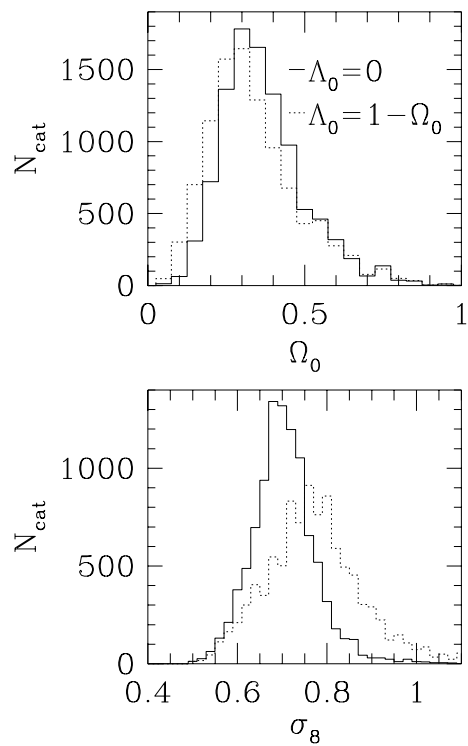}}
\vspace{-1.0cm}
\caption{Histograms showing the distributions of the $10^4$ bootstrap catalogue
best-fitting $\Omega_0$ and $\sigma_8$ values. Open and flat models are 
represented by solid and dotted lines respectively.}
\label{fig:bfhist}
\end{figure}

In order to assess the statistical uncertainties in the estimated parameters, 
one can consider both contours of $\Delta \chi^2$ and the distributions of the
best-fitting values of the bootstrap catalogues. It is interesting to
look at both of these because the $\chi^2$ method formally assumes that
the uncertainty in the estimated temperature function is Gaussian
distributed, which is not fully justified given the small number of
clusters in the two samples. Fig.~\ref{fig:bfhist} shows the distribution
of best-fitting bootstrap catalogue parameters. The open and flat models
are represented by solid and dotted lines respectively.
\begin{figure}
\centering
\vspace{-1.0cm}
\centerline{\epsfxsize=9.0cm \epsfbox{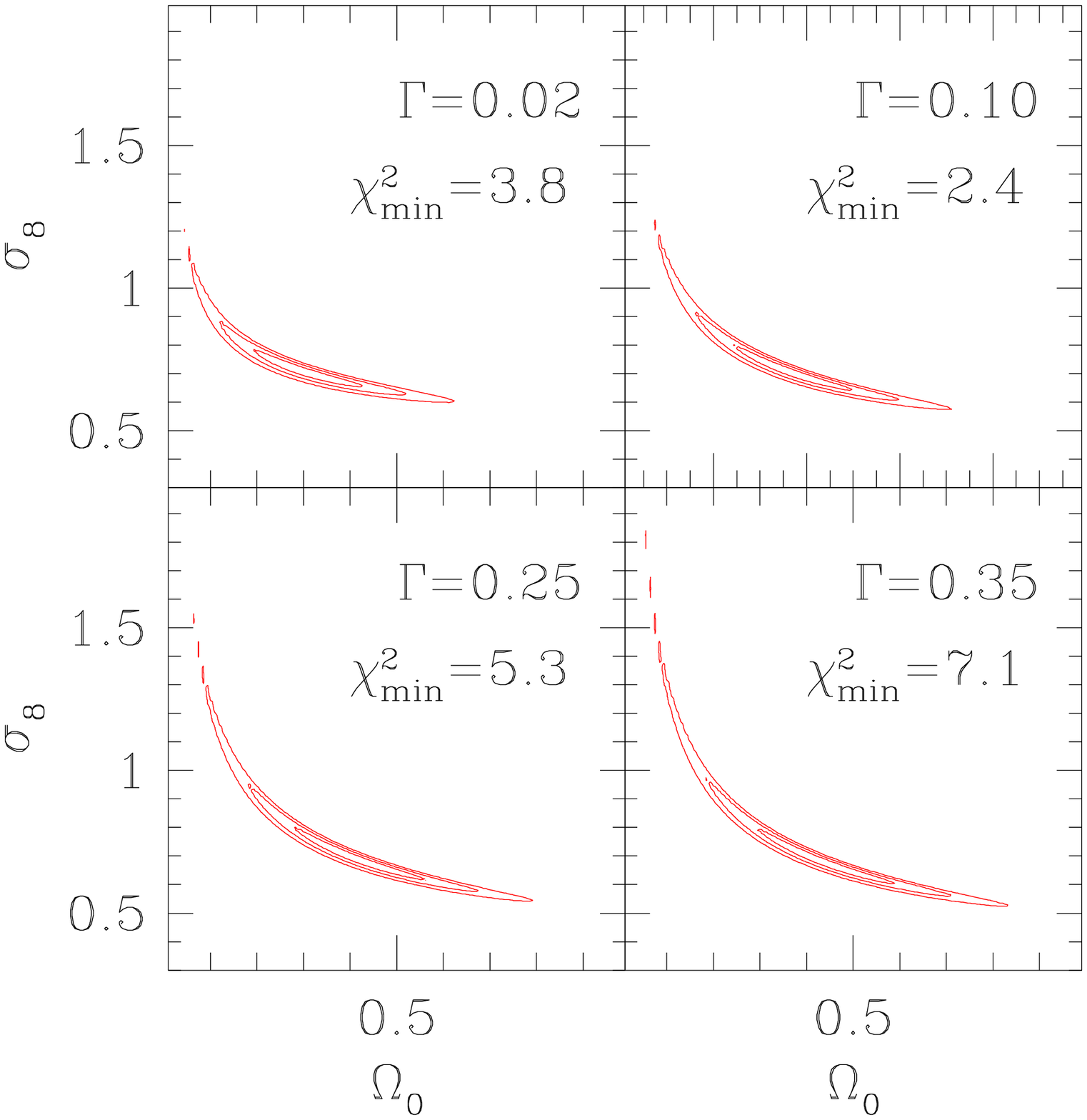}}
\vspace{-2.0cm}
\caption{Contours of constant $\Delta \chi^2$ for slices in parameter space 
at fixed $\Gamma$, as given in the individual panels. The $\Delta \chi^2$
values shown are $2.30$, $6.17$ and $11.8$ and, for two interesting parameters,
these correspond to enclosed probabilities of $1$, $2$ and $3\sigma$.
All panels are for the $\Lambda_0=0$ models. In each case, the minimum $\chi^2$
from all of the models with the chosen $\Gamma$ is also given.}
\label{fig:manychisq}
\end{figure}
\begin{figure}
\centering
\vspace{-1.0cm}
\centerline{\epsfxsize=9.0cm \epsfbox{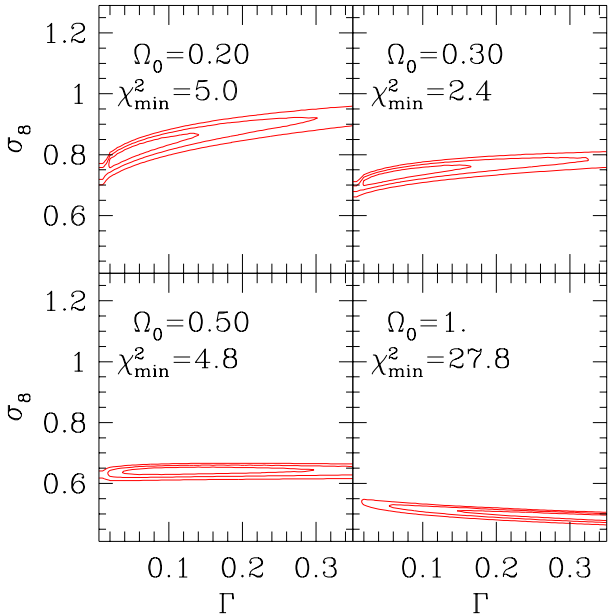}}
\vspace{-2.0cm}
\caption{Contours of constant $\Delta \chi^2$ for slices in parameter space 
at fixed $\Omega_0$, as given in the individual panels. The $\Delta \chi^2$
values shown are $2.30$, $6.17$ and $11.8$ and, for two interesting parameters,
these correspond to enclosed probabilities of $1$, $2$ and $3\sigma$.
All panels are for the $\Lambda_0=0$ models. In each case, the minimum $\chi^2$
from all of the models with the chosen $\Omega_0$ is also given.}
\label{fig:manychisq2}
\end{figure}
Figures~\ref{fig:manychisq} and \ref{fig:manychisq2} show slices through the
three-dimensional parameter space at different values of $\Gamma$ and 
$\Omega_0$ respectively. The contour levels represent $1$, $2$ and $3\sigma$
contours for two degrees of freedom, \ie not marginalised with respect to
the parameter that is fixed. All
panels are for open models with $\Lambda_0=0$. Very little changes with the
inclusion of a non-zero $\Lambda_0$. The minimum values of $\chi^2$ are given
for each of the slices.
Estimates of the uncertainties have been obtained by looking at the
cumulative distributions of the bootstrap catalogue best-fitting parameters.
The following ranges contain $68$ per cent of these values:
\begin{equation}
\Omega_0=[0.24,0.49], \Gamma=[0.02,0.16], \sigma_8=[0.64,0.76]
\label{bfopen}
\end{equation}
if $\Lambda_0=0$, and
\begin{equation}
\Omega_0=[0.20,0.49], \Gamma=[0.02,0.16], \sigma_8=[0.67,0.88]
\label{bflam}
\end{equation}
if $\Lambda_0=1-\Omega_0$. When the individual bootstrap catalogue best-fitting
parameters were overplotted on the $\Delta \chi^2$ contours shown in
Figs.~\ref{fig:manychisq} and~\ref{fig:manychisq2}, they were found to trace
very similar patterns. Only a few lay outside the `$3\sigma$' contours, having
$\Omega_0$ values slightly larger than would be expected given these two 
figures. However, as is evident 
from Fig.~\ref{fig:bfhist}, there are nevertheless
very few catalogues favouring $\Omega_0>0.9$.
This provides some reassurance that the $\Delta \chi^2$ contours
are not providing a misleading impression of the statistical uncertainties in
the various parameters, despite the low number of clusters per bin.

\section{Maximum likelihood parameter estimation}\label{sec:maxl}

One drawback of the method presented in the previous Section is that
the precise cluster redshifts are not employed and so the modelling ignores
any evolution within the redshift ranges spanned by the two data sets.
In this Section, details are 
given of a maximum likelihood technique that has been 
applied to determine the most likely model for producing the measured 
values of cluster temperatures and redshifts. This has the benefit over the 
previous method that no particular binning in temperature is required and 
that the
individual cluster redshifts are used. The disadvantage is that the measured
cluster fluxes are not included in the scheme, and additional assumptions
about the cluster luminosities need to be made.

\begin{figure*}
\centering
\vspace{-4.0cm}
\centerline{\epsfxsize=15.0cm \epsfbox{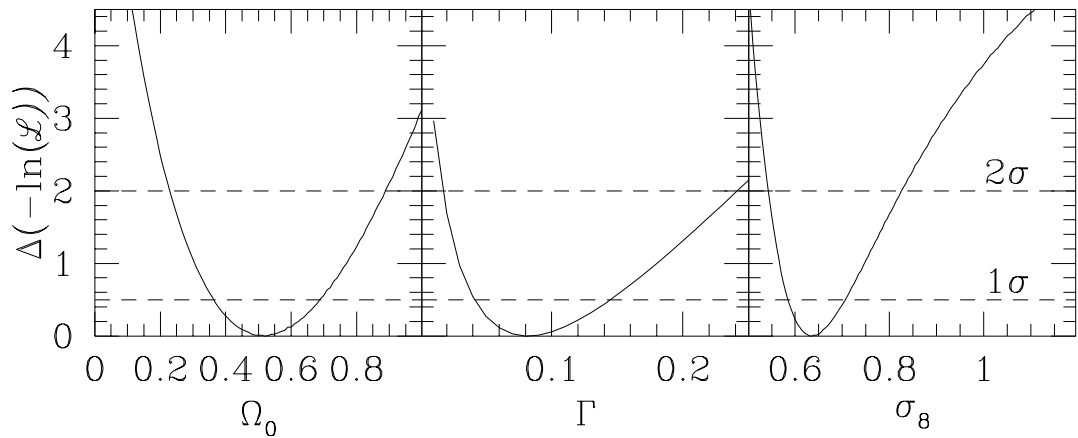}}
\vspace{-3.0cm}
\caption{$\Delta (-$ln likelihood$)$s for $\Omega_0$ (left panel), $\Gamma$
(middle) and $\sigma_8$ (right), marginalised over the other two parameters
in each case. Dashed lines at $1$ and $2\sigma$ significance are shown, and 
the top of each panel corresponds to $3\sigma$ for one interesting parameter.
The default assumptions giving rise to these results are described in the 
text.}
\label{fig:like1}
\end{figure*}
If $a(z,T) \d z\d T$ is the model probability of observing a cluster
with temperature $T$ at redshift $z$, then the likelihood of the observed
distribution of cluster temperatures and redshifts can be written as
\begin{equation}
{\cal L}=\Pi_{{\rm i}=1}^{\rm N}a_{\rm i} \d z\d T e^{-a_{\rm i} \d z\d T}
\Pi_{{\rm j}={\rm N}+1}^{\rm N_{Tot}} e^{-a_{\rm j} \d z\d T}.
\label{like0}
\end{equation}
Here j is an index running over all those of the ${\rm N_{Tot}}$ total
bins in redshift and temperature that are empty, and i runs
over just the N bins occupied by the observed clusters.
This equation follows from assuming that the
cluster redshifts and temperatures are independent, and taking bins such that
$a_{\rm i}\ll 1 ~\forall {\rm i}$. Neglecting the terms that do not
depend upon the cosmological parameters,
equation~(\ref{like0}) can be rewritten as
\begin{equation}
{\rm ln} {\cal L}= \Sigma_{{\rm i}=1}^{\rm N} a_{\rm i} -\int\int a \d z \d T.
\label{like}
\end{equation}
The function $a(z,T)$ can be defined as
\begin{equation}
a(z,T)=n(z,T,p)\frac{\d V(z,p)}{\d z}\zeta(z,T,p)
\label{defna}
\end{equation}
where $p$ represents the three model parameters being investigated,
$n$ is the comoving cluster number density 
given by the Press-Schechter-based model including a
correction for the $20$ per cent uncertainty in the 
mass-temperature conversion, 
$V$ is the comoving volume and $\zeta$ is the product of the fraction of sky 
surveyed and the estimated completeness of the survey. For the high
redshift catalogue, the fraction of sky surveyed has been calculated by
integrating over all possible fluxes at a given $(z,T,p)$, thereby accounting
for the scatter in the luminosity-temperature relation and the variation of
sky covered as a function of flux in the {\it EMSS}. A single temperature
bremsstrahlung model has been used to calculate the `K'-corrections to the
cluster fluxes. The default 
temperature to luminosity conversions that have been 
employed are given by equation~(\ref{ltmean01}) for the $0.3-3.5$~keV band and,
again using the HA91 clusters,
\begin{equation}
\log_{10}\bar{L}_{37}^{2-10}=3.93\log(kT/{\rm keV})-2.92
\label{ltmean21}
\end{equation}
for the $2-10$~keV band. The scatter in log$_{10}($luminosity$)$ about these 
relations is described by equation~(\ref{ltscat01}) and by 
\begin{equation}
\sigma=0.70-0.52\log_{10}(kT/{\rm keV})
\label{ltscat02}
\end{equation}
respectively.
To estimate how the choice of $L-T$ conversion affects the results,
the David \etal (1993) sample has also been considered, giving
\begin{equation}
\log_{10}\bar{L}_{37}^{0.3-3.5}=2.76\log(kT/{\rm keV})-2.02
\label{ltmean02}
\end{equation}
and
\begin{equation}
\log_{10}\bar{L}_{37}^{2-10}=3.45\log(kT/{\rm keV})-2.57.
\label{ltmean22}
\end{equation}
As in the previous Section, there is assumed to be no evolution in the
luminosity-temperature relation.

The dependence of the best-fitting parameters on the following factors has
been investigated: the assumed completeness of the low-redshift sample, the
assumed conversion between cluster flux and {\it EMSS} detect cell flux, the
luminosity-temperature conversions, and the amplitudes of the scatter about
these mean relations. The default choice is to assume 
that both high and low-redshift samples are $100$ per cent complete.
Note that a lower completeness for the 
high-redshift sample would give rise to lower estimates for $\Omega_0$.
Unless stated otherwise, a cluster at redshift $z$ is
taken to have $2.4-z$ times as much flux as would fall into the {\it EMSS} 
detect cell (see figure~$1$ of Henry \etal 1992). 
The redshift and temperature limits for the binning are chosen
so as just to encompass all of the data points. For the low-redshift sample
this means $2.5<$kT/keV$<10$ and $0<z<0.1$, and for the high-redshift bins
$3.7<$kT/keV$<10$ and $0.3<z<0.4$. The region of parameter space
that has been searched for the most likely parameters is $0.01\le\Omega_0\le1$
(with $\Lambda_0=0$ or $1-\Omega_0$), $0.01\le\Gamma\le0.25$ and 
$0.5\le\sigma_8\le1.2$.

\subsection{Results}\label{ssec:mlresults}

\begin{figure}
\centering
\centerline{\epsfxsize=12.0cm \epsfbox{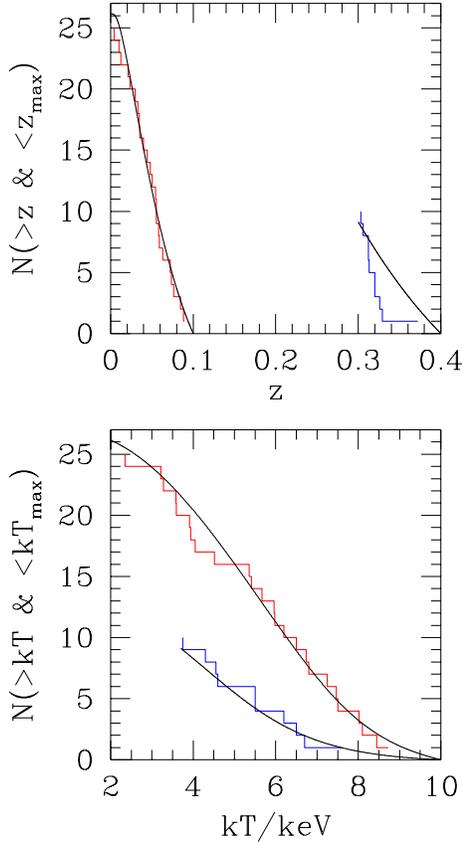}}
\caption{Marginalised data (stepped curves) and best-fitting model (smooth 
curves) for the redshift and temperature distributions of the clusters in the
best-fitting model: $\Omega_0=0.52$ $(\Lambda_0=0)$, $\Gamma=0.08$ and
$\sigma_8=0.63$.}
\label{fig:like2}
\end{figure}

With the default assumptions described in the preceeding section,
the most likely parameter values are $\Omega_0=0.52$, $\Gamma=0.08$ and
$\sigma_8=0.63$ for the $\Lambda_0=0$ model. The statistical uncertainties 
on these parameters can be determined by considering the marginalised
likelihoods. These one-dimensional likelihoods are shown for all three of
the fitted parameters in Fig.~\ref{fig:like1}. In each case the maximum
likelihood is found at each value of the chosen parameter, independently of
the other two parameters. Confidence levels are therefore taken for a single
interesting parameter, leading to $1\sigma$
uncertainties of $\Omega_0=0.52^{+0.17}
_{-0.16}$, $\Gamma=0.08^{+0.07}_{-0.04}$ and $\sigma_8=0.63^{+0.08}
_{-0.05}$. The model predictions for
the marginalised number of clusters as a function of either redshift or
temperature are shown in Fig.~\ref{fig:like2} for the most likely case.
It is evident that the high-redshift sample cluster distribution is not
very well matched by the model. In fact a Kolmogorov-Smirnov test indicates
that the probability that this data set was produced from the model distribution
is of order $10^{-3}$. A comparison of {\it CNOC} (Carlberg, Yee \&
Ellingson 1997a) with {\it EMSS} (Henry \etal 1992) redshifts
suggests that the low probability cannot be accounted for 
by redshift measurement errors.
If the maximum redshift limit of the Henry (1997) cluster sample
is decreased from $0.4$ to $0.33$ and the likelihoods are recomputed then the
resulting $\Omega_0$ becomes $\lsim 0.13$ (the best-fitting $\sigma_8$ is at
least $1.2$, which is the largest value considered on the grid of likelihoods).
This suppression of the high-z survey
volume with the corresponding removal of only one of the $10$ clusters has 
the effect of reducing the apparent evolution and, likewise $\Omega_0$. 
However, if the redshift distribution of the $z>0.4$ {\it EMSS} clusters is 
also considered, then models with so little evolution will overproduce
distant objects so vastly that even the additional 
uncertainties in such predictions are unlikely to account for the results. 
The redshift distribution of the $0.3<z<0.4$ clusters is therefore rather
difficult to explain.
\begin{table}
\centering
\begin{center}
\caption{Most likely parameters for the different input assumptions shown
in Fig.~\ref{fig:like3}.}
\begin{tabular}{llll} \hline
Change from default choices & $\Omega_0$ & $\Gamma$ & $\sigma_8$ \\
\hline
none & 0.52 & 0.08 & 0.63 \\
$\beta_{\rm TM}=1.2$ & 0.55 & 0.09 & 0.68 \\
$\beta_{\rm TM}=1.5$ & 0.61 & 0.10 & 0.73 \\
$\beta_{\rm TM}=2.0$ & 0.72 & 0.11 & 0.81 \\
$f_{\rm clus}/f_{\rm det}=2.0-z$ & 0.59 & 0.07 & 0.61 \\
$f_{\rm clus}/f_{\rm det}=2.8-z$ & 0.44 & 0.10 & 0.67 \\
$z=0$ sample $90$ per cent complete & 0.58 & 0.08 & 0.62 \\
$z=0$ $L-T$ scatter $\times 0.7$ & 0.59 & 0.11 & 0.61 \\
$z=0.3-0.4$ $L-T$ scatter $\times 1.3$ & 0.56 & 0.07 & 0.62 \\
$L-T$ given by eqns.~(\ref{ltmean02}) and~(\ref{ltmean22}) & 0.43 & 0.05 & 
0.67 \\
$z=0.3-0.4$ mean $L|T \times 1.5$ & 0.64 & 0.06 & 0.60 \\
$\Lambda_0=1-\Omega_0$ & 0.45 & 0.10 & 0.69 \\
\hline
\hline
\end{tabular}
\end{center}
\end{table}

\begin{figure*}
\centering
\centerline{\epsfxsize=15.0cm \epsfbox{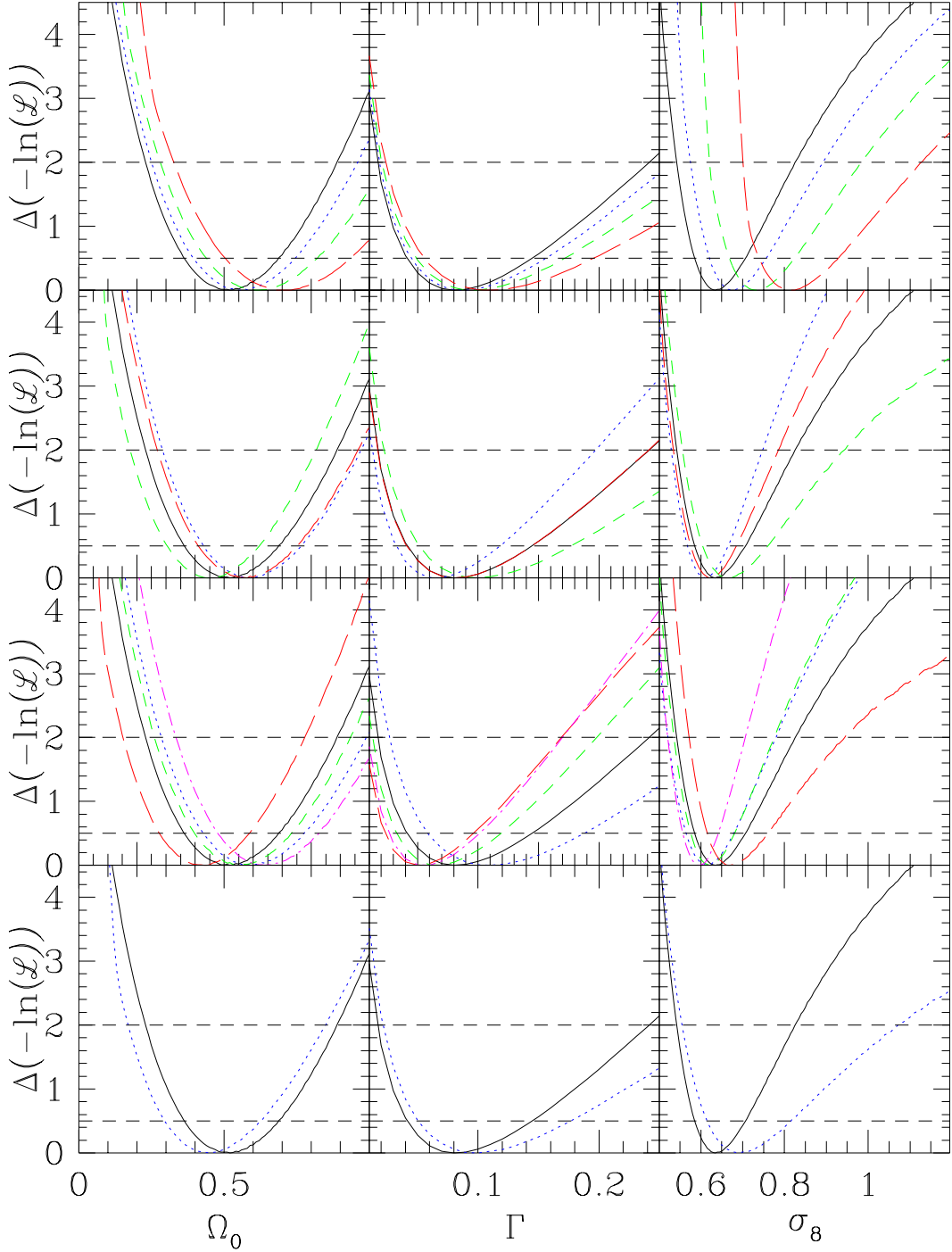}}
\caption{$\Delta (-$ln likelihood$)$s for $\Omega_0$ (left panels), $\Gamma$
(middle) and $\sigma_8$ (right), marginalised over the other two parameters
in each case. Dashed lines at $1$ and $2\sigma$ significance are shown, and 
the top of each panel corresponds to $3\sigma$ for one interesting parameter.
In all rows, the default curves shown in Fig.~\ref{fig:like1} are reproduced
with solid lines.
The top row shows curves for $\beta_{\rm TM}=1.2, 1.5$ and $2$ with dotted,
short dashed and long dashed lines respectively. In the second row, the results
of a cluster to {\it EMSS} detect cell flux ratio of $2-z$ (dotted) and $2.8-z$
(short dashed) are given along with results obtained by assuming that the
low-redshift data are $10$ per cent incomplete (long dashed). 
Illustrated in the third row are curves resulting from
decreasing the $L-T$ scatter for the low-z sample by $30$ per cent (dotted),
increasing the scatter by $30$ per cent for the high-z sample (short dashed),
using the different mean $L-T$ relations described in 
equations~(\ref{ltmean22}) and~(\ref{ltmean02}) (long dashed) or simply
increasing the high-z luminosity at a given temperature by $50$ per cent
(dot-dashed). The final row shows the 
likelihoods found after assuming that $\Lambda_0=1-\Omega_0$ (dotted).}
\label{fig:like3}
\end{figure*}

Fig.~\ref{fig:like3} shows the effect on the most likely parameters 
produced by altering the model assumptions. From top to
bottom, the rows show the variation with $1$) $\beta_{\rm TM}$, $2$) 
survey incompleteness and uncertain $f_{\rm clus}/f_{\rm det}$, $3$) $L-T$
treatment and $4$) the inclusion of a non-zero $\Lambda_0$ term. Table $2$ 
contains details of the most likely parameters in the different cases.

Increasing $\beta_{\rm TM}$ has the effect of boosting $\sigma_8$. This
suppresses the amount of evolution at any given $\Omega_0$, thus leading
to a higher estimate for the density parameter. 
It should be noted that a much more 
effective way of changing the most likely $\Omega_0$ would be to have some {\it
evolution} in the value of $\beta_{\rm TM}$, rather than merely changing the
mean at all redshifts. 
Even $\beta_{\rm TM}=2$, which seems rather extreme (see
Section~\ref{sec:omnotone}), only produces a $+0.2$ change in the density
parameter.

The second row in Fig.~\ref{fig:like3} shows the relative insensitivity of
the results to the assumed conversion between total cluster flux and {\it EMSS}
detect cell flux, $\Delta \Omega_0 = \pm0.08$ for 
extreme choices (see Henry \etal 1992 figure~$1$).
When the ratio $f_{\rm clus}/f_{\rm det}$ is taken to be larger, this 
corresponds to a need for more luminous clusters in the high-redshift sample
and hence a lower value for $\Omega_0$.
A $10$ per cent incompleteness in the low-redshift cluster sample enhances the
evolution enough to create $\Omega_0=0.58$. However, if the high-redshift
sample was less than $100$ per cent complete then the
best-fitting value would decrease accordingly
(see Section~\ref{sec:omnotone}).

The scatter in the luminosity-temperature relations has been varied 
to see how much the most likely density parameter can be increased. In the
third row of Fig.~\ref{fig:like3}, it is apparent that very little impact is
made with these alterations. More significant is the reduction in $\Omega_0$
that comes from exchanging the $L-T$ equations~(\ref{ltmean02}) 
and~(\ref{ltmean22}) for the default choices~(\ref{ltmean01}) 
and~(\ref{ltmean21}). This yields $\Omega_0=0.43$. The amplification of
the high-redshift luminosity at a given temperature by $50$ per cent, which is 
more than is observed (Tsuru \etal 1996; Mushotzky \& Scharf 1997; Henry 1997),
increases the number of clusters that should be seen further away. 
Consequently, to fit the observed abundance, it is necessary to increase the
amount of evolution to compensate, giving a best-fitting $\Omega_0=0.64$.

As is evident from this figure, the best-fitting value of $\Omega_0$ is 
difficult to change by more than $\sim 0.1$ with any single one of these
systematic changes. Additional reasons why these results might be 
misleading will
be considered in Section~\ref{sec:omnotone}, with particular emphasis on
what can be said about the health of the critical density model in the light
of cluster evolution.

The final row in Fig.~\ref{fig:like3} shows how the most likely parameters
are effected by the inclusion of a $\Lambda_0=1-\Omega_0$ term. 
In this case, the most likely parameters and $1\sigma$ marginalised
uncertainties are $\Omega_0=0.45^{+0.18}_{-0.16}$, 
$\Gamma=0.10^{+0.08}_{-0.05}$ and $\sigma_8=0.69^{+0.13}_{-0.08}$. 

Only the choice of upper redshift limit for the Henry (1997) sample and the 
excessive change in $\beta_{\rm TM}$ of the systematic
effects investigated above, manage to alter the most likely $\Omega_0$ by much
more than $0.15$. However, the statistical uncertainties are sufficiently large
that in none of the above cases is $\Omega_0=1$ ruled out with a significance
of $3\sigma$ or greater. 

\section{Comparison with other studies}\label{sec:others}

The analysis presented here produces results in good agreement with those of
Henry (1997). This is not surprising, given that very nearly identical data 
have been employed, albeit within different modelling procedures. These
results fit happily with the {\it ROSAT} logN-logS analysis by Mathiesen \&
Evrard (1998) as long as the luminosity evolves like $L\propto (1+z)^{\sim 3}$.
This is approximately what is implied by equations~(\ref{kev})
and~(\ref{ltmean01}), although it should be noted that the results in 
Section~\ref{sec:evol} are almost completely insensitive to the 
assumed form of the L-T relation, unlike those in Section~\ref{sec:maxl}.
Kitayama \& Suto (1997) have also used the logN-logS relation of X-ray 
clusters to arrive at the conclusion that $\Omega_0=0.45$ is favoured if
$\Lambda_0=0$.

Of the other results in this field, there appear to be two more distinct camps,
one on either side of the above settlement in the middle of $\Omega_0$-space.
On the lower-side, although not by much, favouring 
$\Omega_0\approx0.40\pm0.12$ and $0.30\pm0.10$ are Carlberg \etal 
(1997b) and Bahcall \etal (1997) respectively. These results are consistent
with those found here, although the small quoted uncertainties strongly
imply that $\Omega_0 < 1$. The `high-ground', with 
$\Omega_0=0.8-1$, is occupied by Sadat \etal (1998) and 
Blanchard \& Bartlett (1997). At this point, the reader might wonder why 
any of these conclusions concerning $\Omega_0$ are worth taking
seriously, given that they appear to be inconsistent and span the whole
range of interesting values. That such disparate results are obtained is
presumably a suggestion that unaccounted-for systematic errors are lurking
beneath the surface in at least one of these analyses. As has been stressed 
in this paper, and was also discussed by Pen (1998), the
cluster number density evolution is rapid if considered at a particular mass.
However, at a fixed number density the mass varies relatively little over the
range of redshifts probed by the presently available data. For instance, in
an $\Omega_0=1$ model normalised to match the $z=0$ cluster abundance, the
evolution between $z=0.05$ and $z=0.33$ of the mass at a constant 
comoving cluster number density is $\sim 45$ per cent. Any 
unaccounted-for relative systematic temperature errors (\ie affecting the 
low- and high-redshift samples differently) must therefore be smaller than 
$\sim 30$ per cent, or else the comparison between the models and 
observations will become completely 
unreliable. The quantification
of any potential systematic uncertainties in mass estimation is thus
crucial in determining the size of the error bar on the derived $\Omega_0$.

For the studies producing
lower-$\Omega_0$ determinations, velocity dispersions obtained using
different procedures are used to estimate masses. Frenk \etal (1996) and 
van Haarlem, Frenk \& White (1997) 
investigated the biases in mass estimation using velocity dispersions, and
more recently Borgani \etal (1997) considered the model dependence of these
biases. These studies suggest that only after very careful modelling
of the velocity dispersion calculation procedure, should this method be applied
with any confidence. From the simple scaling laws $M\propto \sigma_{\rm v}^3$ 
and $M\propto T^{1.5}$, it is immediately apparent that fractional 
uncertainties in velocity dispersion would need to be half those in the X-ray
temperature in order for the virial mass to be similarly uncertain. 
The importance of such conversions is shown by figures~$1$ and~$2$ of Carlberg
\etal (1997b) where the higher redshift {\it CNOC} number density determination
lies at a greater mass than the other {\it CNOC} point. This is slightly
surprising because the virial theorem suggests that the same velocity
dispersion should be produced by equal masses out to a given physical radius,
independent of redshift. Under the assumption that both {\it CNOC} redshift
ranges have the same cluster mass limit, which was the choice made by 
Bahcall \etal (1997),
the higher redshift {\it CNOC} point can be made to lie rather happily on the
$\Omega_0=1$ model prediction shown in Carlberg et al.s figure~$2$, as long as 
their high-redshift point is shifted to the mass limit of their low-redshift 
sample. While it is not clear that this is the correct mass to choose, and
Bahcall \etal took an intermediate value, this demonstrates how sensitively
the conclusions about $\Omega_0$ depend on such details. 
In addition to these difficulties, the {\it CNOC} sample, which is almost a
subsample of {\it EMSS} clusters and contains $5$ from the high-redshift set
studied by Henry (1997), has both a luminosity and a velocity dispersion cut.
Consequently, when converting to an observed number density of clusters 
above a particular mass, a correction for scatter in the correlations
between these properties is required. As Carlberg \etal (1997b) showed, this
is uncertain by $\sim 50$ per cent.

Bahcall \etal also consider the richnesses in the Palomar Distant Cluster
Survey. This is suspect because of the well-known difficulties 
associated with trying to calibrate the richness-mass correlation
and, more importantly, any possible evolution in it (\eg van Haarlem,
Frenk \& White 1997).

Moving on to the large-$\Omega_0$ camp, Blanchard \& Bartlett claim that 
their result ``in conjunction with the absence of any negative evolution in 
the luminosity-temperature relation, provides robust evidence in favour of a
critical density universe''. The data used to draw this conclusion, come
from a variety of sources: {\it CNOC} velocity dispersions, {\it EMSS}
luminosities and {\it ROSAT} luminosities. As these authors point out, there
is considerable uncertainty in the number density of clusters above $6$~keV
at redshift zero. 
A consistent value of $\sim 30\times10^{-8} (\Mpc)^{-3}$ is obtained
from both the samples of Edge \etal (1990) and 
HA91. Integrating the {\it ROSAT} 
Brightest Cluster Sample (BCS) (Ebeling \etal 1997) best-fitting Schechter
function for the {\it EMSS} passband ($0.3-3.5$~keV) from infinity to the
luminosity corresponding to $6$~keV ($1.25\times10^{37}h^{-2}$W) suggests
a number density of clusters of $\sim 40\times10^{-8} (\Mpc)^{-3}$.
The value calculated by Blanchard \& Bartlett is twice as large as this. 
They derived their estimate by integrating in the {\it ROSAT}
passband to a limit determined from the L-T relation of Arnaud \& Evrard
(in preparation). The consequence of having a larger redshift zero abundance
is to increase the apparent evolution and hence increase the derived $\Omega_0$.

Sadat \etal (1998) used the lack of evolution 
of the cluster luminosity-temperature relation to give a ``tentative''
measurement of $\Omega_0=0.85\pm0.2$. This determination was based on an
application of the method proposed by Oukbir \& Blanchard (1997). They
showed how, by using the redshift zero cluster temperature function 
normalisation and the redshift distribution of {\it EMSS} clusters, it was
possible to relate $\Omega_0$ to the evolution of the cluster
luminosity-temperature relationship. However, as Sadat \etal note, Oukbir
\& Blanchard (1997) used the uncorrected (see Eke \etal 1996) HA91
temperature function, which gives an overabundance of present-day clusters.
This will feed through into the $\Omega_0$ determination (equation~$7$ of 
Sadat et al.) and produce an
overestimate, although it is not clear by how much. The measurement of the
L-T evolution found by Sadat \etal relies upon the correct 
choice of the slope for this relation. It is important because the more 
distant observed clusters are intrinsically hotter and more
luminous than the nearby ones, and if the slope of the $L-T$ relation
(they choose $L\propto T^3$) is not accurate then some spurious evolution (or 
lack of evolution) would be introduced because of the different types of
clusters being probed at high and low redshifts. For the
low-redshift clusters contained in their table~$1$ the mean temperature is 
$6.05$ keV and the mean luminosity
is $15.9\times10^{37}W$. From table~$2$ in their paper, the mean temperature
and luminosity for clusters satisfying $0.4<z<0.6$ are $7.13$ keV and
$55\times10^{37}W$. If the $L-T$ relation is non-evolving then this suggests
that $L\propto T^{\sim 7}$ (This exponent is still $>6$ if the one 
high-redshift cooling flow cluster RXJ1347.5-1145 is removed from this 
calculation.), so there is an apparent inconsistency in the method
that is likely to have an impact on the derived value of $\Omega_0$.
Their figure~$3$ and equation~$5$
suggest that, if the high-z clusters are biased to have large luminosities,
then this would cause $\Omega_0$ to be overestimated.

\section{Order statistics and the EMSS cluster sample}\label{sec:order}

A number of recent papers have reported the existence of 
large mass concentrations at 
redshifts above $0.5$ (Luppino \& Gioia 1995, Donahue 1996,
Luppino \& Kaiser 1997, Donahue \etal 1998).
Very often, these findings are accompanied by a claim that the existence of 
such objects proves troublesome for hierarchically clustering models of
structure formation, particularly if $\Omega_0=1$. 
However, the magnitude of this difficulty is not quantified. The 
purpose of this Section is to propose and then apply a method by which the 
extent of the problem can be gauged.
To this end, the question: `What is the expected redshift of
the nth furthest galaxy cluster with a temperature above $kT$?', 
will be addressed. The answer will depend sensitively
upon $\Omega_0$ because of the dependence of the growth
rate on $\Omega_0$. Thus, one should be able to use the
observations of hot, distant clusters to discriminate between different
values of $\Omega_0$. More specifically, given that any observational
survey is likely to be less than $100$ per cent complete, the presence of a
hot, massive cluster can be used to 
set an upper limit on $\Omega_0$. 

\subsection{Method and application}\label{ssec:how}

The Press-Schechter expression for the abundance of haloes as a function of
mass has been tested using
large N-body simulations out to about $z=0.5$ for haloes having
masses corresponding to $kT\sim 5$ keV in an $\Omega_0=1$ universe (\eg Eke \etal 1996). 
It appears to provide a good description of the cluster mass
function over this range of redshifts when the spherical collapse threshold
density is used. The more recent, and larger,
simulation of Tozzi \& Governato (1997) suggests that 
the Press-Schechter equation overestimates the observed evolution
in very massive haloes. However, it is not yet clear how much of this
apparent discrepancy results from their particular choice of groupfinder.
Whilst the theoretical expectation is still somewhat uncertain for 
very rare objects at $z\gsim0.5$, this is an issue that can 
be addressed in the near future with very large-volume dark matter 
simulations (\eg Evrard \etal in preparation), so this need not be an
insurmountable barrier to the usefulness of the approach described here. 

As was described in Section~\ref{sec:model}, the simple
spherical collapse model will be used in conjunction with the Press-Schechter
expression in order to provide a description of the halo mass function.
Additionally, $\bar{\beta}_{\rm TM}=1$ will be assumed to hold at all 
redshifts, along with a $20$ per cent scatter about this mean value.
Using the results of Eke \etal (1996), the normalisation of the mass
fluctuation spectrum will be taken as $0.50\pm0.04$ for the $\Omega_0=1$
model considered in this Section. (Note that this is slightly different 
from the best-fitting value of 0.52 found by Eke \etal (1996). This new 
estimate takes into account both the systematic
overestimation of $\sigma_8$ resulting from the measured temperature
uncertainties (see table~$1$ of Eke et al.) and another, slightly larger, term 
(neglected in their table) arising from scatter in the mass-temperature 
relation.)

From these stipulations, it is possible to calculate the 
redshift distribution of all clusters hotter than a specified value. 
The probability that, from the
population of clusters hotter than a particular temperature, a cluster is
situated at a redshift no greater than $z$ will be denoted by 
$P(\leq z) \equiv P(\leq z~|~kT>kT_{\rm min})$. This
quantity is calculable from the model for any given temperature. 
For brevity and clarity the temperature dependence is suppressed in the
following expressions.
Using this
notation the probability that the furthest cluster (at $z_1$) has a redshift
greater than z can be written
\begin{equation}
P(z_1>z) = 1-P(z_1 \leq z) = 1-(P(\leq z))^{N},
\label{simple}
\end{equation}
where $N$ is the total number of clusters above the chosen 
temperature limit in the whole volume, given by the Press-Schechter formula. 
For a particular redshift distribution, larger $N$ will probe 
further into the tail of the distribution and $z_1$ will be expected to be 
larger.
This procedure can be extended to give the redshift distributions of the $k$th
furthest cluster in the following manner. The probability
that the $k$th furthest cluster (hotter than $kT_{\rm min}$) has a redshift 
greater than a particular value is given by
\begin{equation}
P(z_k>z) = 1 - P(z_k \leq z).
\label{simple2}
\end{equation}
To calculate the last term in this equation, it is
necessary to consider
the various combinations of clusters that would give rise to $z_k \leq z$.
This probability can then be written, using the combination function $C$
\footnote{The combination function is $^N C_j \equiv N!/(j!(N-j)!)$.}, as
\begin{equation}
P(z_k \leq z) = \sum_{j=0}^{k-1}~^{N}C_j~P(\leq z)^{N-j} (1 - P(\leq z))^j
\label{notsimple}
\end{equation}
and this, together with equation~(\ref{simple2}) gives $P(z_k > k)$.
This probability will be very sensitive to both the model normalisation
$\sigma_8$ and the error in the measured cluster temperatures. Consequently,
a marginalisation technique allowing for Gaussian errors on both of these
quantities has been employed.
This involves integrating over all normalisations and temperatures, weighting
with the probabilities that these are in fact the actual values of $\sigma_8$ 
and the lowest temperature of the clusters under consideration.
\begin{figure}
\centering
\centerline{\epsfxsize=8.0cm \epsfbox{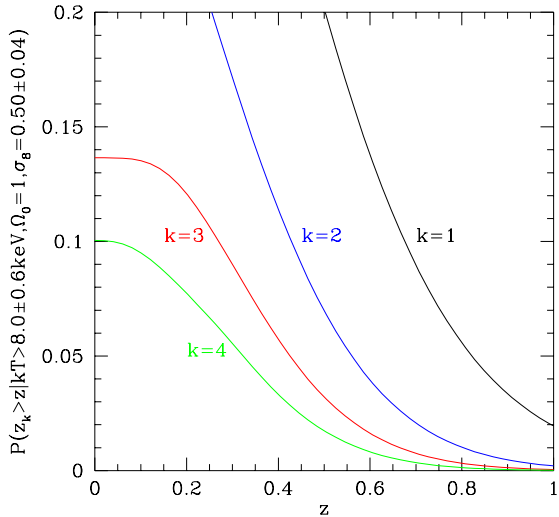}}
\caption{The probability that the $k$th furthest of all {\it EMSS} clusters
with a temperature exceeding $8.0\pm0.6$~keV has a redshift larger than $z$,
assuming a cosmological model with $\Omega_0=1$, $\sigma_8=0.50\pm0.04$ and
$\Gamma=0.25$. There is less than a $1$ in $7$ chance that three such
objects will exist in an {\it EMSS} cluster survey area of sky, and only a 
probability of $\sim 0.02$ that the third furthest hot cluster will have a 
redshift greater than $0.53$.}
\label{fig:mix}
\end{figure}

The three clusters that will be used in this analysis are $MS0015.9+1609$ at
$z=0.54$, $MS0451.6-0305$ at $z=0.54$ and $MS1054.5-0321$ at $z=0.83$. 
{\it ASCA}
temperatures of $8.0\pm0.6$~keV (Mushotzky \& Scharf 1997), $10.4\pm1.2$~keV
(Donahue 1996) and $14.7\pm2.1$~keV (Donahue \etal 1998) respectively have 
been found for these objects. Therefore in the {\it EMSS} cluster sample,
the third furthest cluster having a temperature greater than or equal to 
$8.0\pm0.6$~keV has a redshift larger than $0.53$. Given that these clusters
were all found during a search containing only $735$ square degrees of sky, 
this value, rather than the entire $4\pi$ steradians will be used for 
calculating the model predictions.

\subsection{Results}\label{ssec:what}

As can be seen in Figure~\ref{fig:mix}, the probability that three clusters
with $kT \geq 8.0\pm0.6$~keV exist in an $\Omega_0=1$, $\sigma_8=0.50\pm0.04$,
$\Gamma=0.25$ model is less than $0.14$. Furthermore, it is another factor of
$\sim 6$ less likely that the third most distant of these has a redshift of 
more than $0.53$, as is observed. The conclusion from this exercise would
appear to be that such a 
model can be ruled out at the $98$ per cent confidence level. This would
improve to $\sim 99$ per cent if an extra {\it EMSS} cluster hotter than
$8$~keV were found at a redshift larger than $0.53$.
It is worth
reiterating that this result would vary somewhat if the Press-Schechter
treatment of the halo mass function was badly awry, or if the best-fitting
$\sigma_8$ were underestimated because of a poor choice of the factor
relating mass to temperature, $\beta_{\rm TM}$. Therefore, it would be
premature to consider the existence of these hot clusters at high redshifts
as being sufficient to rule out $\Omega_0=1$ cosmologies. Performing the
above analysis for an $\Omega_0=0.52$ and $\sigma_8=0.63\pm0.06$ model
(keeping $\Gamma$ fixed at $0.25$) yielded a probability of $0.38$ for the 
third furthest of the {\it EMSS} $kT \geq 8.0\pm0.6$~keV clusters having
$z>0.53$.
\begin{table*}
\centering
\begin{center}
\caption{The main strengths and weaknesses of the three methods used here 
for constraining $\Omega_0$ using cluster evolution.}
\begin{tabular}{lll} \hline
Method&Strengths&Weaknesses\\
\hline
$\chi^2$ (Section~\ref{sec:evol})& Utilises the cluster flux measurements. & 
Requires carefully selected data.\\
& Almost independent of $L-T$ relation. & Results depend slightly on $z$ and 
$T$ bin choice.\\
&&\\
M.L. (Section~\ref{sec:maxl})& All cluster redshifts are used. & Requires 
carefully selected data.\\
& No $z$ or $T$ bin choices are necessary. & Uncertainty introduced with 
$L-T$ relation.\\
&&\\
Order statistics (Section~\ref{sec:order})& No need for a statistically 
well-defined data sample. & Based on a model extrapolation. \\
& Independent of $L-T$ relation. & Results are very sensitive to the assumed 
$\sigma_8$.\\
\hline
\end{tabular}
\end{center}
\end{table*}

\section{Overview of the different techniques}\label{sec:tech}

Having analysed cluster data in three distinct ways, it is worth discussing
the strengths and weaknesses of the different methods that
were employed in Sections~\ref{sec:evol},~\ref{sec:maxl} 
and~\ref{sec:order}. While all of these techniques use similar input
data in an attempt to determine $\Omega_0$, they are sufficiently
diverse to merit some discussion. Table~1 contains a summary of
the main pros and cons of these approaches.

The main drawback with the first two methods, which is not so important for
the order statistics approach, is the need for a statistically well-defined 
sample of data with well-understood selection effects. Given such 
observational catalogues, the particular data that are used by the $\chi^2$
and maximum likelihood (M.L.) techniques differ somewhat. In the former
case, only the cluster temperatures and fluxes are important. The 
$1/V_{\rm max}$ estimator enables this method to sidestep the need for any
cluster luminosity-temperature relation, apart from a minor incompleteness
correction. However, the redundancy of the individual redshift measurements,
other than to determine if a cluster should be included in the `high-' or 
`low-'z temperature function, leaves a certain freedom to choose what redshifts
to assign to the fits. The choice of temperature bins is also something that
can alter the resulting $\Omega_0$ determination to a small extent. In 
contrast, the M.L. method uses all the cluster temperatures and redshifts and
avoids awkward binning problems. However, the limits of integration in 
equation~(\ref{like}), coming from the adopted ranges in $z$ and $T$, do create
some freedom in the most likely parameters. This comes as a result both of the
peculiar redshift distribution of the high-z clusters, and the uncertainties
in the selection functions at the extreme temperatures and redshifts.
Additionally, the price paid for not using the measured flux values
is that one must adopt a luminosity-temperature relation
(at all relevant redshifts) and any error in the assumed form of this
relation will introduce a systematic error into the estimated $\Omega_0$.

The difference between the derived best-fitting parameters from the $\chi^2$
and M.L. techniques
has non-negligible contributions from the systematic uncertainties mentioned
in the previous paragraph. Looking at the situation more positively, it is
perhaps reassuring that, despite the various differences between the two
methods, such similar results were returned.

An opportunity to win the jackpot is offered by the order statistics method;
namely the chance to rule out $\Omega_0=1$ with observations of only a few 
very hot and distant clusters. However, the size of the payout for an
apparently successful data set is not yet well-defined because of the
uncertainties in both the cluster mass function evolution and the 
mass-temperature conversion for such rare objects at such high redshifts. 
Nevertheless, these are both eminently measureable using numerical simulations,
and the Hubble volume simulation performed by the Virgo consortium 
(Evrard \etal in preparation) 
should address the former of these issues in the near future. It is
worth noting that the potential non-virialisation of large mass 
concentrations at high redshifts, or the boosting in emission-weighted
temperature coming from a recent merger event should not cause any difficulties
for this method. This is because such effects are included in hydrodynamical 
simulations and would be calibrated through the mean $\beta_{\rm TM}$ and
the scatter around this value.

\section{What went wrong?}\label{sec:omnotone}

Given that all `right-minded' cosmologists {\it know} that $\Omega_0=1$, a
pertinent question to ask at this point, on their behalf, is 
``What has gone
wrong?'', \ie what would it need for an $\Omega_0=1$ model to be more 
favoured by cluster evolution.
In this Section, consideration is given to the magnitudes of the errors that
are required in some of the preceeding assumptions.

If $\Omega_0$ is in fact unity,
then it is very probable that 
at least one of the observations or the model used above
is wrong. The first of these possibilities seems unlikely. 
An $\Omega_0=1$ model would
require a systematic overestimation of the high-redshift cluster temperatures 
by {\it ASCA} of about $25$ per cent, or a corresponding underestimation of the
low-redshift temperatures by this same amount. This corresponds to a shift in
all of
the cluster temperatures that is typically twice the quoted uncertainty and,
as such, seems highly unlikely. Having said that, the analysis of {\it ASCA}
data by Markevitch \etal (1998) showed, for single temperature fits to $15$
of the original $25$ HA91 clusters, an average $6$ per cent
increase in the measured temperatures. While this is not sufficiently large,
if such a systematic shift were applied to the entire low-redshift temperature
function, then the most likely $\Omega_0$ would increase by $\sim 20$ per cent.

Contamination of the X-ray emission by active galactic nuclei (AGN)
or cooling flows
would alter the mass-temperature conversion.
The effect of AGN contamination is
to increase the emission-weighted temperature. Cooling flows
have the opposite effect. These potential complications are better addressed
from the point of view of decontaminating the data, rather than 
adding extra processes into
the numerical simulations, so will be considered as `observational'
problems. The likelihood of any interloping AGN in the
high-redshift sample is very low, and high-resolution images show no evidence
for such contamination.
More important is the effect of cooling flows on the two different redshift 
samples. If cooling flows were more prevalent in the low-redshift
sample than the high-redshift one, then their removal would result 
in an increased separation of the two temperature functions,
thus favouring higher $\Omega_0$. By fitting both single-temperature
and two isothermal models to {\it ASCA} data, Markevitch \etal (1998) showed
that, for the $15$ low-z clusters in common with HA91, the decontaminated
emission-weighted temperature is, on average, $6$ per cent larger than the
single-temperature value.
For the most discrepant single cluster, the effect was $5$ times as great.
Allen (1998) performs a similar analysis including two of the Henry (1997)
(\ie high-z) clusters, and finds the temperatures increase by $15$ and $24$
per cent when a cool component is removed. There are two simple reasons 
why one might expect
cooling flows to be more prominent in the {\it EMSS} high-z sample rather
than the HA91 catalogue. Firstly the densities are larger and hence 
cooling times are shorter in the higher redshift range. 
Secondly, the {\it EMSS} selection algorithm
tends to favour clusters that have centrally peaked emission, 
with only those exceeding a flux threshold in a $2'.4 \times 2'.4$ 
detect cell being included in the catalogue. 
If the results from Allen (1998) are
typical of the whole Henry (1997) sample, then the effect of cooling
flow decontamination would be to increase the high-redshift cluster 
temperatures by more than the low-redshift ones. This
would somewhat overcompensate for the systematic temperature shift
described in the previous paragraph, leading to an overall estimate of 
$\Omega_0$ that is about $20$ per cent lower than the value found in 
Section~\ref{sec:maxl}, where these considerations were neglected.

As was
discussed in Section~\ref{sec:maxl}, a $10$ per cent incompleteness in the
low-redshift sample only increases $\Omega_0$ by $0.06$, so this is not a
sufficiently large effect to help the critical density model. 
Independent checks are available on the
sample completenesses. These come from the comparison of luminosity functions
obtained from the cluster catalogues used here
with those from the larger {\it ROSAT} cluster samples at low 
(Ebeling \etal 1997) and high (Rosati \etal 1998) redshifts. 
{\it ROSAT} does not detect enough very luminous clusters to compare with the 
majority of the luminosity range of the {\it EMSS} high-z sample,
because of the smaller fraction of sky it surveyed. Nevertheless, where 
an overlap in luminosity exists, 
there is agreement between the cluster abundances
in the different surveys. These comparisons have enough statistical 
uncertainty to make them reassuring rather than highly restrictive. There
could still plausibly be $\sim 40$ per cent difference between the various 
samples. In order to favour $\Omega_0=1$ though, there would need to be a
significantly greater incompleteness in the low-redshift sample than exists
in the more distant one, and this seems unlikely.

The considerations in the preceeding few paragraphs suggest that if anything,
relative to the results found in Sections~\ref{sec:evol} and~\ref{sec:maxl},
lower rather than higher values of $\Omega_0$ would be favoured by the 
potential `observational' systematic effects. Thus, if the data used in this 
paper are the product of an $\Omega_0=1$ universe then it seems that
the finger of suspicion should 
point firmly in the direction of the modelling of cluster evolution.

\begin{figure}
\centering
\vspace{-2.0cm}
\centerline{\epsfxsize=8.5cm \epsfbox{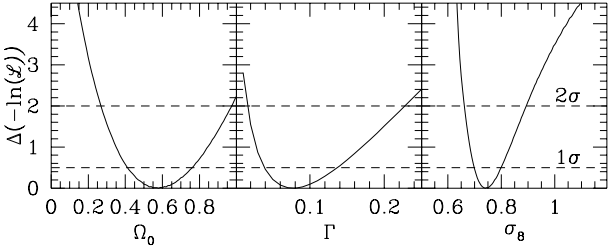}}
\vspace{-2.0cm}
\caption{$\Delta (-$ln likelihood$)$ for $\Omega_0$ (left panel), $\Gamma$
(middle) and $\sigma_8$ (right), marginalised over the other two parameters
in each case. Dashed lines at $1$ and $2\sigma$ significance are shown, and 
the top of each panel corresponds to $3\sigma$ for one interesting parameter.
The curves are calculated assuming that the collapse threshold is given by
$\delta_{\rm eff}=1.48(1+z)^{-0.06}$ rather than the spherical collapse
$\delta_{\rm c}$.}
\label{fig:delteff}
\end{figure}

\begin{figure}
\centering
\vspace{-2.0cm}
\centerline{\epsfxsize=8.5cm \epsfbox{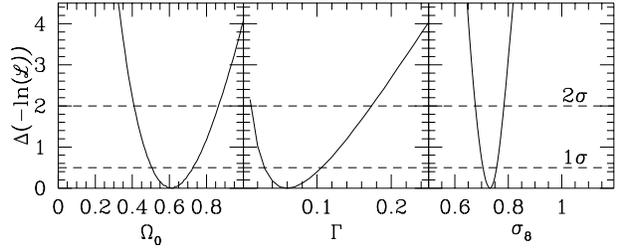}}
\vspace{-2.0cm}
\caption{As for Fig.~\ref{fig:delteff}, but with the redshift information
for the $9~z>0.4$ {\it EMSS} clusters included in the likelihood
calculation.}
\label{fig:delteff2}
\end{figure}

\begin{figure}
\centering
\centerline{\epsfxsize=12.0cm \epsfbox{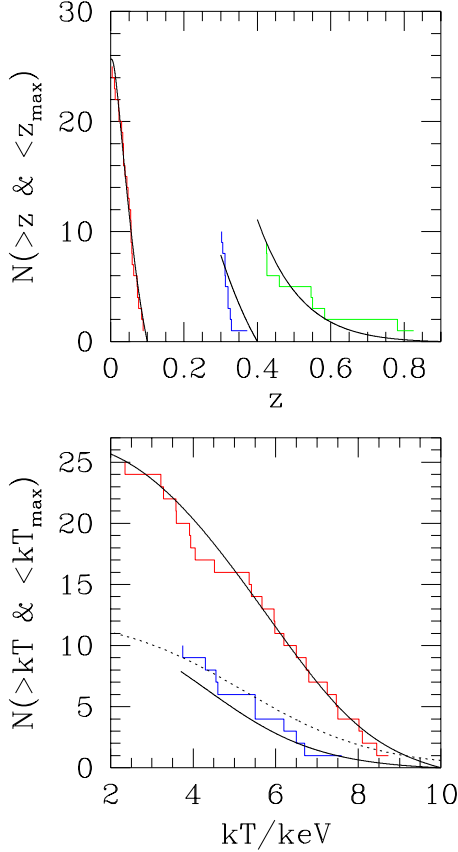}}
\caption{Marginalised distributions of clusters. Calculated assuming that
the threshold varies with redshift according to $\delta_{\rm eff}$, and
using the $9~z>0.4$ extra {\it EMSS} clusters. Smooth curves are shown for the 
most likely model, namely $\Omega_0=0.62$, $\Gamma=0.06$ and $\sigma_8=0.53$. 
The dotted
curve in the temperature panel corresponds to the model prediction for the 
distribution of the temperatures that are yet to be measured. The stepped 
curves represent the observed cluster distributions.}
\label{fig:delteff3}
\end{figure}

While the statement that the Press-Schechter expression provides a good
description of the mass function of galaxy-cluster-sized objects had stood
up to much scrutiny, the recent report by Tozzi \& Governato (1997)
suggests that it overestimates the actual evolution for $\Omega_0=1$. Their
simulation is of a cube with side $500 \Mpc$ and thus allows the evolution of
very rare objects to be traced back to higher redshifts than was possible
with the smaller volumes used in previous studies. These authors suggest that
the spherical collapse threshold $\delta_{\rm c}$ should be replaced by
$\delta_{\rm eff}=1.48(1+z)^{-0.06}$. This result appears to be
slightly discrepant with previous findings where comparison is possible, but
it may be that this is just an artefact of the groupfinder that was employed
(see Cole \& Lacey 1996 for more details). Nevertheless, it is 
worthwhile to investigate the possibility that an evolving threshold is
required. Assuming that $\delta_{\rm eff}$ is the form of 
$\delta_{\rm c}$ that is applicable
to all $\Lambda_0=0$ models (similar redshift evolution is found in a large
$\Omega_0=0.3$ CDM simulation; Governato, private communication), 
the effect of such a varying threshold on the
most likely value of $\Omega_0$ can be investigated using the likelihood
procedure described in Section~\ref{sec:maxl}. The resulting marginalised
likelihoods are shown in Fig.~\ref{fig:delteff}. 
It is apparent that, whilst
the most likely value of $\Omega_0$ has only increased to $0.58$, the
$\Delta(-$ln${\cal L})$ required to reach $\Omega_0=1$ is now just $2.2$,
\ie only $\sim 2\sigma$. Whilst this gives the impression that a small hit
has been scored on the good ship $\Omega_0<1$, the captain might retaliate
by pointing out that the {\it EMSS} survey found clusters with $z>0.4$ and,
given that the Tozzi \& Governato $\delta_{\rm eff}$ fitted their simulated
mass function out to a redshift of one, these distant objects can now be
included in the likelihood calculation. Temperatures have yet to be 
measured for all of these clusters, but they do have known redshifts, so an
additional contribution to the likelihood is computed by replacing the $a$
in equation~(\ref{like}) by the integral of $a$ over all cluster temperatures
($2-20$~keV in practice) and just comparing the probability of finding
clusters at the observed redshifts. Additionally, the value of 
$f_{\rm clus}/f_{\rm det}$ is taken to be $1.75$, independent of redshift
for the $z>0.4$ {\it EMSS} clusters.
The $9$\footnote{Recent HRI data show that
MS$1610.4+6616$ is a point source (see Henry 1998) and it
is not included.} identified {\it EMSS} clusters with
$z>0.4$ and their corresponding redshifts are taken from Gioia \& Luppino
(1994) and Luppino \& Gioia (1995),
and the marginalised likelihoods resulting from their inclusion are shown in
Fig.~\ref{fig:delteff2}. It should be noted that there are also $9$
unidentified {\it EMSS} sources (Gioia \& Luppino 1994), 
some of which might contribute to this list, and consequently reduce
the estimated value of $\Omega_0$ from that found here. 
The most likely parameters change only slightly, becoming $\Omega_0=0.62$,
$\Gamma=0.06$ and $\sigma_8=0.53$. Furthermore, the
inclusion of the more distant clusters significantly decreases the sizes
of the statistical uncertainties on the most likely parameters, 
such that $\Omega_0=1$ is now excluded at a higher confidence level than
was the case with the original default assumptions, 
despite the favourably evolving $\delta_{\rm eff}$.
Distributions of the numbers of clusters marginalised with respect to
either redshift or temperature are shown in Fig.~\ref{fig:delteff3}.
As the temperatures
have yet to be measured for the highest redshift clusters, this model
`prediction' is shown as a dotted line.

In addition to any uncertainties in the accuracy of the Press-Schechter 
description of the mass function, there is some leeway in the conversion 
from mass to temperature that may 
effect the most likely $\Omega_0$. There are at 
least $5$ options for tinkering in an attempt to salvage $\Omega_0=1$: 
\begin{flushleft}
1. Evolution of the scatter in both the cluster mass-temperature conversion 
and the observational temperatures would give rise to a systematic error in
the estimation of $\Omega_0$. It has been assumed that the intrinsic scatter 
is unchanging and the low- and high-redshift temperature errors are similar
(which the quoted values suggest they are). 
If these scatters at high redshift implied
that a Gaussian smoothing width of $0.4T$ rather than $0.2T$ was appropriate, then
even the $\Omega_0=1$ model cluster temperature function would appear not to 
evolve. However, for the mass-temperature relation, the
low-$\Omega_0$ simulations of Eke \etal (1998) show no significant evolution
in the scatter over this range of redshifts (see Fig.~\ref{fig:beta}). The 
observational uncertainties are relatively small and, only slightly larger
for the high-$z$ sample. The conclusion is therefore that such an
escape for $\Omega_0=1$ is not easily arranged.\\
2. Evolution of $\mu$, the mean molecular weight of the intracluster gas,
has been ignored in the modelling. If it were $25$ per cent larger at $z=0.33$
than $z=0$ then the model predictions for $\Omega_0=1$ would show no evolution
over this redshift range. This change in $\mu$ is both extreme and in a 
counter-intuitive direction.
Mushotzky \& Loewenstein (1997) recently used {\it ASCA} data to show
that the iron abundance in the intracluster medium (ICM) does not evolve over
the range $0 < z \leq 0.3$. Also,
Kauffmann \& Charlot (1998) adopted a semi-analytical approach to the formation
and evolution of elliptical galaxies, concluding that the metallicity of the
ICM is approximately constant out to $z>1$. 
This approach for saving $\Omega_0=1$ can be ruled out.\\
3. Evolution of $\beta_{\rm TM}$ to the extent described for $\mu$ in 
point 2, would
also make $\Omega_0=1$ compatible with the observations. The definition of 
$\beta_{\rm TM}$ 
is such that it would need to be $25$ per cent {\it smaller} at the
larger redshift. As shown by the simulations of Eke \etal (1998) (see 
Fig.~\ref{fig:beta}) this amount of evolution is difficult to justify and, 
if anything, the $\beta_{\rm TM}$ at $z=0.33$
is a few per cent larger than that at low redshift. For the various
cosmologies studied numerically by Bryan \& Norman (1998), a similar lack of 
evolution was also found (see their figure~$4$). 
This suggests that such a 
route will not assist $\Omega_0=1$ models either, and that the most likely
value for $\Omega_0$ quoted here may be slightly biased to a value that is
too large.\\
4. One clear deficiency of the hydrodynamical simulations that have been used 
to relate virial mass to emission-weighted X-ray temperature is the lack of 
any heating of the intracluster gas by supernovae. 
That this is a problem becomes apparent when the simulated cluster
luminosity-temperature relation is compared with the observed one. Whilst the
models and scaling laws show $L\propto T^2$, the observations suggest a steeper
dependence of luminosity on temperature (see equation~(\ref{ltmean01})). 
Allen \& Fabian (1998) find that decontaminating the cooling flow
contribution to the X-ray emission can leave $L\propto T^{\sim 2}$. However,
a similar study by Markevitch (1998) implies that, in order to produce this
shallow slope, some feedback is still 
necessary after the cooling flow correction has been applied. NFW
showed that by preheating gas in the manner suggested by Evrard \& Henry
(1991), the low-temperature simulated clusters could be warmed up enough, 
along with the corresponding decrease in luminosity as a result of the less 
dense cores, to produce a sufficiently steep luminosity-temperature relation
to match observations. To obtain a very crude idea of the maximum
extent to which such a change will affect
the evolution of the $\Omega_0=1$ temperature functions, a simple mapping of
temperature has been applied to the no-preheating value, $(kT)_0$, to give a
`feedback' temperature, $(kT)_1$. The mapping that has been adopted is
\begin{equation}
{\rm log}_{10}(kT)_1=0.57{\rm log}_{10}(kT)_0+0.34.
\label{feedmap}
\end{equation}
\begin{figure}
\centering
\centerline{\epsfxsize=9.0cm \epsfbox{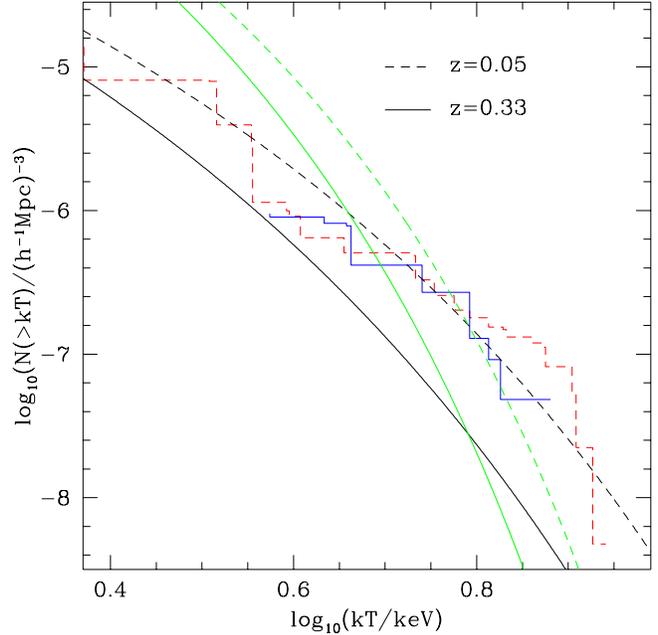}}
\caption{The temperature functions for an
$\Omega_0=1$, $\sigma_8=0.52$ and $\Gamma=0.25$ model are shown with solid and
dashed lines for $z=0.05$ and $z=0.33$ respectively. 
In each case, the curves resulting
from applying the temperature mapping of equation~(\ref{feedmap}) are also
shown. These are steeper than the original predictions. Stepped curves show
the observationally determined temperature functions.}
\label{fig:feed}
\end{figure}
This route for converting the simulated clusters to lie on the observed
luminosity-temperature relation is extreme in its effect on the cluster
temperatures. In practice, one would expect any energy injection to heat the
gas {\it and} decrease the central density. As a result, the simulated
clusters would fall onto the observed $L_{\rm x}-T_{\rm x}$ relation more as
a result of decreasing luminosity than increasing temperature as is assumed
for the mapping above.
Figure~\ref{fig:feed} shows the predictions before and after `feedback'
when the same distortion is applied to both high- and low-redshift
models. With the `feedback' essentially just heating the cooler clusters, it 
produces steeper
cumulative temperature functions, thus favouring lower $\Gamma$ values. 
(It is worth noting that, if the lower temperature clusters were preferentially
contaminated by cooling flows then the estimated $\Gamma$ would be biased low,
but Markevitch (1998) reports that slightly the reverse is seen in his sample
of nearby clusters.) Interestingly, the extreme
feedback simulations of Metzler \& Evrard (1994) increased the gas temperature
of a $5.6$~keV 
cluster by $\sim 15$ per cent. This size of fractional change is 
only reached with the above mapping for clusters cooler than $\sim 4.5$~keV.
Despite all of this, the predicted amount of evolution still exceeds that 
observed. Only if such a prescription were proscribed for the high-redshift 
sample alone would feedback be
effective in reducing the amount of evolution seen in $\Omega_0=1$ models.\\
5. The effects of having a two-phase intracluster gas medium or magnetic
fields are also lacking from the hydrodynamical simulations used to normalise
the mass-temperature relation. While it seems unlikely that they should have
a different effect on the two redshift ranges considered here, they could still
bias the best-fitting parameters as was illustrated in Section~\ref{sec:maxl},
by shifting a different part of the mass function into the observed range of
temperatures. The amount by which $\beta_{\rm TM}$ can be affected is roughly
constrained 
by comparisons between weak lensing masses and X-ray masses. Present
results (Smail \etal 1997; Allen 1998) appear to imply that this additional
physics is not sufficiently misleading to allow $\Omega_0=1$ to be consistent
with the lack of evolution of the cluster number density.
\end{flushleft}

\section{Conclusions}\label{sec:conc}

In this paper, the value of the density parameter
$\Omega_0$ has been constrained using three
different, yet similar, methods. The results are entirely consistent with
those presented by Henry (1997). Combining the results from 
Sections~\ref{sec:evol} and~\ref{sec:maxl} gives the most likely
parameters as $\Omega_0\approx0.44\pm0.2$, 
$\Gamma\approx0.08\pm0.07$ and $\sigma_8\approx0.67\pm0.1$ if
$\Lambda_0=0$. The inclusion of a non-zero cosmological constant term has very
little effect, leading to $\Omega_0\approx0.38\pm0.2$, 
$\Gamma\approx0.09\pm0.08$ and $\sigma_8\approx0.74\pm0.1$. 
Apart from the difficulty arising because of the redshift distribution of
the Henry (1997) sample, none of the systematic effects that have
been considered alter the best-fitting $\Omega_0$ by more than one of the
statistical standard deviations quoted above.
Unless an important systematic error
or uncertainty has been overlooked in the analysis, these results would 
appear to be limited by a lack of data, so, while this preliminary study 
suggests
that the $\Omega_0=1$ model is still bobbing about unhappily in the water 
at a rejection level of $2-3\sigma$, a larger sample of clusters would offer
the opportunity, with this method alone, to sink it properly. Enough ammunition
should be provided by the upcoming X-ray telescope missions {\it XMM} and
{\it AXAF}.

\section*{ACKNOWLEDGMENTS}

Doug Burke and Peter Coles are thanked for helpful discussions. George
Efstathiou is thanked for providing a speedy computer. 
VRE, SMC and CSF acknowledge the support of PPARC postdoctoral,
advanced and senior fellowships. JPH was supported by NASA grants 
NAG 5-2523 and NAG 5-4828.

\end{document}